%% file: paper-rev.tex
\documentclass[11pt]{article}
\usepackage{psfig}
\def\Tu{\hat T}
\def\Td{\check T}

\def\xu{\hat x}
\def\xd{\check x}
\def\xb{\bar x}

\def\Var{\mbox{Var}}

\def\beq{\begin{eqnarray}}
\def\eeq{\end{eqnarray}}

\def\T{{\widetilde T}}

\begin{document}

\fontsize{11}{14.5pt}\selectfont

\begin{center}

{\small Technical Report No.\ 9805,
 Department of Statistics, University of Toronto}

\vspace*{0.5in}

{\LARGE \bf Annealed Importance Sampling} \\[16pt]

{\Large Radford M. Neal}\\[3pt]
 Department of Statistics and Department of Computer Science \\
 University of Toronto, Toronto, Ontario, Canada \\
 \texttt{http://www.cs.utoronto.ca/$\sim$radford/} \\
 \texttt{radford@stat.utoronto.ca}\\[10pt]

 \begin{tabular}{rl} First version: & 18 February 1998 \\
                     Revised:       & 1 September 1998
 \end{tabular}

\end{center}

\vspace{8pt} 

\noindent \textbf{Abstract.} Simulated annealing --- moving from a
tractable distribution to a distribution of interest via a sequence of
intermediate distributions --- has traditionally been used as an
inexact method of handling isolated modes in Markov chain samplers.
Here, it is shown how one can use the Markov chain transitions for
such an annealing sequence to define an importance sampler.  The
Markov chain aspect allows this method to perform acceptably even for
high-dimensional problems, where finding good importance sampling
distributions would otherwise be very difficult, while the use of
importance weights ensures that the estimates found converge to the
correct values as the number of annealing runs increases. This
annealed importance sampling procedure resembles the second half of
the previously-studied tempered transitions, and can be seen as a
generalization of a recently-proposed variant of sequential importance
sampling.  It is also related to thermodynamic integration methods for
estimating ratios of normalizing constants.  Annealed importance
sampling is most attractive when isolated modes are present, or when
estimates of normalizing constants are required, but it may also be
more generally useful, since its independent sampling allows one to
bypass some of the problems of assessing convergence and
autocorrelation in Markov chain samplers.

\section{Introduction}\vspace*{-10pt}

In Bayesian statistics and statistical physics, expectations of
various quantities with respect to complex distributions must often be
computed.  For simple distributions, we can estimate expectations by
sample averages based on points drawn independently from the
distribution of interest.  This simple Monte Carlo approach cannot be
used when the distribution is too complex to allow easy generation of
independent points.  We might instead generate independent points from
some simpler approximating distribution, and then use an importance
sampling estimate, in which the points are weighted to compensate for
use of the wrong distribution.  Alternatively, we could use a sample
of dependent points obtained by simulating a Markov chain that
converges to the correct distribution.  I show in this paper how these
two approaches can be combined, by using an importance sampling
distribution defined by a series of Markov chains.  

This method is inspired by the idea of ``annealing'' as a way of
coping with isolated modes, which leads me to call it \emph{annealed
importance sampling}.  The method is especially suitable when
multimodality may be a problem, but may be attractive even when it is
not, since it allows one to bypass some of the problems of convergence
assessment.  Annealed importance sampling also supplies an estimate
for the normalizing constant of the distribution sampled from.  In
statistical physics, minus the log of the normalizing constant for a
canonical distribution is known as the ``free energy'', and its
estimation is a long-standing problem.  In independent work, Jarzynski
(1997a,b) has described a method primarily aimed at free energy
estimation that is essentially the same as the annealed importance
sampling method described here.  I will focus instead on statistical
applications, and will discuss use of the method for estimating
expectations of functions of state, as well as the normalizing
constant.

Importance sampling works as follows (see, for example, Geweke 1989).
Suppose that we are interested in a distribution for some quantity,
$x$, with probabilities or probability densities that are proportional
to the function $f(x)$.  Suppose also that computing $f(x)$ for any
$x$ is feasible, but that we are not able to directly sample from the
distribution it defines.  However, we are able to sample from some
other distribution that approximates the one defined by $f(x)$, whose
probabilities or probability densities are proportional to the
function $g(x)$, which we are also able to evaluate.

We base our estimates on a sample of $N$ independent points,
$x^{(1)},\ldots,x^{(N)}$, generated from the distribution defined by
$g(x)$.  For each $x^{(i)}$, we compute an importance weight as 
follows:\vspace*{-8pt}
\beq
  w^{(i)} & = & f(x^{(i)}) \ \big/\ g(x^{(i)})
\label{eq-impw}
\eeq
We can then estimate the expectation of $a(x)$ with respect
to the distribution defined by $f(x)$ by\vspace*{-8pt}
\beq
  \bar a & = & \sum_{i=1}^N w^{(i)} a(x^{(i)}) 
               \ \Big/\ \sum\limits_{i=1}^N w^{(i)}
\label{eq-is}
\eeq
Provided $g(x) \ne 0$ whenever $f(x) \ne 0$, it is easy to see that
$N^{-1} \sum w^{(i)}$ will converge as $N\rightarrow\infty$ to
$Z_f/Z_g$, where $Z_f=\int\! f(x)\, dx$ and $Z_g=\int\! g(x)\, dx$ are
the normalizing constants for $f(x)$ and $g(x)$.  One can also see
that $\bar a$ will converge to the expectation of $a(x)$ with respect to 
the distribution defined by $f(x)$.  

The accuracy of $\bar a$ depends on the variability of the importance
weights.  When these weights vary widely, the estimate will
effectively be based on only the few points with the largest weights.
For importance sampling to work well, the distribution defined by
$g(x)$ must therefore be a fairly good approximation to that defined
by $f(x)$, so that the ratio $f(x)/g(x)$ does not vary wildly.  When
$x$ is high-dimensional, and $f(x)$ is complex, and perhaps
multimodal, finding a good importance sampling distribution can be
very difficult, limiting the applicability of the method.

An alternative is to obtain a sample of dependent points by simulating
a Markov chain that converges to the distribution of interest, as in
the Metropolis-Hastings algorithm (Metropolis, {\em et~al} 1953;
Hastings 1970).  Such Markov chain methods have long been used in
statistical physics, and are now widely applied to statistical
problems, as illustrated by the papers in the book edited by Gilks,
Richardson, and Spiegelhalter (1996).

Markov chains used to sample from complex distributions must usually
proceed by making only small changes to the state variables.  This
causes problems when the distribution contains several
widely-separated modes, which are nearly isolated from each other with
respect to these transitions.  Because such a chain will move between
modes only rarely, it will take a long time to reach equilibrium, and
will exhibit high autocorrelations for functions of the state
variables out to long time lags.

The method of simulated annealing was introduced by Kirkpatrick,
Gelatt, and Vecchi (1983) as a way of handling multiple modes in an
optimization context.  It employs a sequence of distributions, with
probabilities or probability densities given by $p_0(x)$ to $p_n(x)$,
in which each $p_j$ differs only slightly from $p_{j+1}$.  The
distribution $p_0$ is the one of interest.  The distribution $p_n$ is
designed so that the Markov chain used to sample from it allows
movement between all regions of the state space.  A traditional scheme
is to set $p_j(x) \propto p_0(x)^{\beta_j}$, for $1 = \beta_0 >
\beta_1 > \cdots > \beta_n$.

An annealing run is started at some initial state, from which we first
simulate a Markov chain designed to converge to $p_n$, for some number
of iterations, which are not necessarily enough to actually approach
equilibrium.  We next simulate some number of iterations of a Markov
chain designed to converge to $p_{n-1}$, starting from the final state
of the previous simulation.  We continue in this fashion, using the
final state of the simulation for $p_j$ as the initial state of the
simulation for $p_{j-1}$, until we finally simulate the chain designed
to converge to $p_0$.

We hope that the distribution of the final state produced by this
process is close to $p_0$.  Note that if $p_0$ contains isolated
modes, simply simulating the Markov chain designed to converge to
$p_0$ starting from some arbitrary point could give very poor results,
as it might become stuck in whatever mode is closest to the starting
point, even if that mode has little of the total probability mass.
The annealing process is a heuristic for avoiding this, by taking
advantage of the freer movement possible under the other
distributions, while gradually approaching the desired $p_0$.
Unfortunately, there is no reason to think that annealing will give
the precisely correct result, in which each mode of $p_0$ is found
with exactly the right probability.  This is of little consequence in
an optimization context, where the final distribution is degenerate
(at the maximum), but it is a serious flaw for the many applications
in statistics and statistical physics that require a sample from a
non-degenerate distribution.

The annealed importance sampling method I present in this paper is
essentially a way of assigning weights to the states found by multiple
simulated annealing runs, so as to produce estimates that converge to
the correct value as the number of runs increases.  This is done by
viewing the annealing process as defining an importance sampling
distribution, as explained below in Section~\ref{sec-ais}.  After
discussing the accuracy of importance sampling in general in
Section~\ref{sec-acc}, I analyse the efficiency of annealed importance
sampling in Section~\ref{sec-eff}, and find that good results can be
obtained by using a sufficient number of interpolating distributions,
provided that these vary smoothly.  Demonstrations on simple
distributions in Section~\ref{sec-demo} and on a statistical problem
in Section~\ref{sec-lin} confirm this. 

Annealed importance sampling is related to tempered transitions (Neal
1996), which are another way of modifying the annealing procedure so
as to produce correct results.  As discussed in Section~\ref{sec-tt},
annealed importance sampling will sometimes be preferable to using
tempered transitions.  When tempered transitions are still used, the
relationship to annealed importance sampling allows one to find
estimates for ratios of normalizing constants that were previously
unavailable.  Section~\ref{sec-sis} shows how one can also view a form
of sequential importance sampling due to MacEachern, Clyde, and Liu
(1998) as an instance of annealed importance sampling.  Finally, in
Section~\ref{sec-disc}, I discuss the general utility of annealed
importance sampling, as a way of handling multimodal distributions, as
a way of calculating normalizing constants, and as a way of combining
the adaptivity of Markov chains with the advantages of independent
sampling.

\section{The annealed importance sampling 
         procedure}\vspace*{-10pt}\label{sec-ais}

Suppose that we wish to find the expectation of some function of $x$
with respect to a distribution with probabilities or probability
densities given by $p_0(x)$.  We have available a sequence of other
distributions, given by $p_1(x)$ up to $p_n(x)$, which we hope will
assist us in sampling from $p_0$, and which satisfy $p_j(x)\ne0$
wherever $p_{j-1}(x)\ne0$.  For each distribution, we must be able to
compute some function $f_j(x)$ that is proportional to $p_j(x)$.  We
must also have some method for sampling from $p_n$, preferably one
that produces independent points. Finally, for each $i$ from $1$ to
$n\!-\!1$, we must be able to simulate some Markov chain transition,
$T_j$, that leaves $p_j$ invariant.

The sequence of distributions used can be specially constructed to
suit the problem, but the following scheme may be generally useful.
We fix $f_0$ to give the distribution of interest, and fix $f_n$ to
give the simple distribution we can sample from, and then let
\beq
  f_j(x) & = & f_0(x)^{\beta_j}\, f_n(x)^{1-\beta_j}
\label{eq-family}
\eeq
where $1=\beta_0 > \beta_1 > \ldots > \beta_n=0$.  Note that the
traditional simulated annealing scheme with $f_j(x) = f_0(x)^{\beta_j}$ 
would usually be less suitable, since it usually
leads to a $p_n$ for which independent sampling is not easy.  

For applications in Bayesian statistics, $f_n$ would be the prior
density, which is often easy to sample from, and $f_0$ would be the
unnormalized posterior distribution (the product of $f_n$ and the
likelihood).  When only posterior expectations are of interest,
neither the prior nor the likelihood need be normalized.  When the
normalizing constant for the posterior (the marginal likelihood) is of
interest, the likelihood must be properly normalized, but the prior
need not be, as discussed below.

The Markov chain transitions are represented by functions
$T_j(x,x^{\prime})$ giving the probability or probability density of
moving to $x^{\prime}$ when the current state is $x$.  It will not be
necessary to actually compute $T_j(x,x^{\prime})$, only to generate an
$x^{\prime}$ from a given $x$ using $T_j$.  These transitions may be
constructed in any of the usual ways (eg, Metropolis or Gibbs sampling
updates), and may involve several scans or other iterations.  For the
annealed importance sampling scheme to be valid, each $T_j$ must leave
the corresponding $p_j$ invariant, but it is not essential that each
$T_j$ produce an ergodic Markov chain (though this would usually be
desirable).

Annealed importance sampling produces a sample of points, $x^{(1)},
\ldots, x^{(N)}$, and corresponding weights, $w^{(1)}, \ldots,
w^{(N)}$.  An estimate for the expectation of some function, $a(x)$,
can then be found as in equation~(\ref{eq-is}).  To generate each point, 
$x^{(i)}$, and associated weight, $w^{(i)}$, we first generate a 
sequence of points, $x_{n-1},\ldots,x_0$, as follows:
\beq\begin{array}{l}
  \mbox{Generate $x_{n-1}$ from $p_n$.} \\[3pt]
  \mbox{Generate $x_{n-2}$ from $x_{n-1}$ using $T_{n-1}$.} \\[3pt]
  \mbox{\hspace*{70pt}\ldots} \\[3pt]
  \mbox{Generate $x_1$ from $x_2$ using $T_2$.} \\[3pt]
  \mbox{Generate $x_0$ from $x_1$ using $T_1$.} 
\end{array}\label{eq-samp}\eeq
We then let $x^{(i)} = x_0$, and set
\beq
  w^{(i)} & = & {f_{n-1}(x_{n-1}) \over f_n(x_{n-1})}\,
                {f_{n-2}(x_{n-2}) \over f_{n-1}(x_{n-2})}\,
                \cdots\,  
                {f_1(x_1) \over f_2(x_1)}\,
                {f_0(x_0) \over f_1(x_0)}
\label{eq-aisw}
\eeq
To avoid overflow problems, it may be best to do the computations 
in terms of $\log(w^{(i)})$.

To see that annealed importance sampling is valid, we can consider an
extended state space, with points $(x_0,\ldots,x_{n-1})$.  We identify
$x_0$ with the original state, so that any function of the
original state can be considered a function of the extended state, by
just looking at only this component.  We define the distribution for
$(x_0,\ldots,x_{n-1})$ by the following function proportional to the
joint probability or probability density:
\beq
  f(x_0,\ldots,x_{n-1}) & = & f_0(x_0)\, \T_1(x_0,x_1)\, \T_2(x_1,x_2)\,
                              \cdots\, \T_{n-1}(x_{n-2},x_{n-1})
\label{eq-extf}
\eeq
Here, $\T_j$ is the reversal of the transition defined by $T_j$.  That is,
\beq
  \T_j(x,x^{\prime}) & = & T_j(x^{\prime},x)\, p_j(x^{\prime})\, /\, p_j(x)
                     \ \ =\ \ T_j(x^{\prime},x)\, f_j(x^{\prime})\, /\, f_j(x)
\label{eq-rev}\eeq
The invariance of $p_j$ with respect to $T_j$ ensures that these are
valid transition probabilities, for which $\int
\T_j(x,x^{\prime})\,dx^{\prime} = 1$.  This in turn guarantees that
the marginal distribution for $x_0$ in (\ref{eq-extf}) is the same as 
the original distribution of interest (since the joint probability
there is the product of this marginal probability for $x_0$ and
the conditional probabilities for each of the later components given the
earlier components).

For use below, we apply equation~(\ref{eq-rev}) to rewrite the function 
$f$ as follows:
\beq
  f(x_0,\ldots,x_{n-1}) & = & f_0(x_0)\, 
   {f_1(x_0) \over f_1(x_0)}\, \T_1(x_0,x_1)\, 
       \cdots\, 
   {f_{n-1}(x_{n-2}) \over f_{n-1}(x_{n-2})}\, \T_{n-1}(x_{n-2},x_{n-1}) \\[3pt]
  & = &
   {f_0(x_0) \over f_1(x_0)}\, T_1(x_1,x_0)\,
   \cdots\,
   {f_{n-2}(x_{n-2}) \over f_{n-1}(x_{n-2})}\, T_{n-1}(x_{n-1},x_{n-2})\,
   f_{n-1}(x_{n-1})\ \ \ \ \ \
\label{eq-extf2}
\eeq

We now look at the joint distribution for $(x_0,\ldots,x_{n-1})$ defined 
by the annealed importance sampling procedure (\ref{eq-samp}).  It is
proportional to the following function:
\beq
  g(x_0,\ldots,x_{n-1}) & = & f_n(x_{n-1})\, T_{n-1}(x_{n-1},x_{n-2})\,
                              \cdots\, T_2(x_2,x_1)\, T_1(x_1,x_0)
\label{eq-extg}
\eeq
We regard this as an importance sampler for the distribution (\ref{eq-extf}) 
on the extended state space.  The appropriate importance weights are found
using equations~(\ref{eq-impw}), (\ref{eq-extf2}), and~(\ref{eq-extg}).  
Dropping the superscript $(i)$ on the right side to simplify notation, 
they are:
\beq
  w^{(i)} & = & {f(x_0,\ldots,x_{n-1}) \over g(x_0,\ldots,x_{n-1})} 
  \ \ =\ \ 
   {f_0(x_0) \over f_1(x_0)}\,
   {f_1(x_1) \over f_2(x_1)}\, 
   \cdots\,
   {f_{n-2}(x_{n-2}) \over f_{n-1}(x_{n-2})}\, 
   {f_{n-1}(x_{n-1}) \over f_n(x_{n-1})}
\eeq
These weights are the same as those of equation~(\ref{eq-aisw}), showing 
that the annealed importance sampling procedure is valid.

The above procedure produces a sample of single independent points
$x^{(i)}$ for use in estimating expectations as in
equation~(\ref{eq-is}).  In practice, better estimates will often be
obtained if we use each such point as the initial state for a Markov
chain that leaves $p_0$ invariant, which we simulate for some
pre-determined number of iterations.  We can then estimate the
expectation of $a(x)$ by the weighted average (using the $w^{(i)}$) of
the simple average of $a$ over the states of this Markov chain.  This
is valid because the expectation of $a(x)$ with respect to $p_0(x)$ is
the same as the expectation with respect to $p_0(x)$ of the average
value of $a$ along a Markov chain that leaves $p_0$ invariant and
which is started in state $x$ (since if the start state has
distribution $p_0$, all later states will also be from $p_0$).

Annealed importance sampling also provides an estimate of the ratio of
the normalizing constants for $f_0$ and $f_n$.  Such normalizing
constants are important in statistical physics and for statistical
problems such as Bayesian model comparison. The normalizing constant
for $f$, as defined by equation~(\ref{eq-extf}), is the same as that
for $f_0$, and the normalizing constant for $g$ in
equation~(\ref{eq-extg}) is the same as that for $f_n$.  The average
of the importance weights, $\sum w^{(i)} / N$, converges to the ratio
of these normalizing constants, $Z_0/Z_n$, where $Z_0 =
\int\!f_0(x)\,dx$ and $Z_n = \int\!f_n(x)\,dx$.  

In a Bayesian application where $f_n$ is proportional to the prior and
$f_0$ is the product of $f_n$ and the likelihood, the ratio $Z_0/Z_n$
will be the marginal likelihood of the model --- that is, the prior
probability or probability density of the observed data.  Note that
the prior need not be normalized, since any constant factors there
will cancel in this ratio, but the likelihood must include all
constant factors for this estimate of the marginal likelihood to be
correct.

The data collected during annealed importance sampling runs from $p_n$
down to $p_0$ can also be used to estimate expectations with respect
to any of the intermediate distributions, $p_j$ for $0<j<n$.  One
simply uses the states, $x_j$, found after application of $T_{j-1}$ in
(\ref{eq-samp}), with weights found by omitting the factors in
equation~(\ref{eq-aisw}) that pertain to later states.  Similarly, one
can estimate the ratio of the normalizing constants for $f_j$ and
$f_n$ by averaging these weights.

Finally, although we would usually prefer to start annealing runs with
a distribution $p_n$ from which we can generate independent points,
annealed importance sampling is still valid even if the points
$x_{n-1}$ generated at the start of each run are not independent.  In
particular, these points could be generated using a Markov chain that
samples from $p_n$.  The annealed importance sampling estimates will
still converge to the correct values, provided the Markov chain used
to sample from $p_n$ is ergodic.

\section{Accuracy of importance sampling 
         estimates}\vspace*{-10pt}\label{sec-acc}

Before discussing annealed importance sampling further, it is
necessary to consider the accuracy of importance sampling estimates in
general.  These results will also be needed for the demonstrations in
Sections~\ref{sec-demo} and~\ref{sec-lin}.  

For reference, here again is the importance sampling estimate, $\bar a$, 
for $E_f[a]$, based on points $x^{(i)}$ drawn independently from the density 
proportional to $g(x)$:
\beq
  \bar a & = & \sum_{i=1}^N w^{(i)} a(x^{(i)}) 
               \ \Big/\ \sum\limits_{i=1}^N w^{(i)}
       \ \ =\ \ N^{-1} \sum_{i=1}^N w^{(i)} a(x^{(i)}) 
               \ \Big/\ N^{-1} \sum\limits_{i=1}^N w^{(i)}\ \ 
\label{eq-is2}\eeq
where \mbox{$w^{(i)}=f(x^{(i)})\,/\,g(x^{(i)})$} are the importance weights.

The accuracy of this importance sampling estimator is discussed by
Geweke (1989).  An estimator of the same form is also used with
regenerative Markov chain methods (Mykland, Tierney, and Yu 1995;
Ripley 1987), where the weights are the lengths of tours between
regeneration points.

In determining the accuracy of this estimator, we can assume without loss 
of generality that the normalizing constant for $g$ is such that 
$E_g[w^{(i)}]=1$, since multiplying all the $w^{(i)}$ by a constant
has no effect on $\bar a$.  We can also assume that $E_f[a] =
E_g[w^{(i)}a(x^{(i)})] = 0$, since adding a constant to $a(x)$ simply
shifts $\bar a$ by that amount, without changing its variance.
For large $N$, the numerator and denominator on the right side of
equation~(\ref{eq-is2}) will converge to their expectations, which on
these assumptions gives
\beq
  \bar a & \!=\! & \Big( E[w^{(i)} a(x^{(i)})] + e_1 \Big) 
               \ \Big/\ \Big( E[w^{(i)}] + e_2 \Big)
       \ =\ { e_1 \over 1+e_2} \ =\ e_1 - e_1 e_2 + \cdots\ \ \ \ \
\eeq
where $e_1$ and $e_2$ are the differences of the averages from their
expectations.  When $N$ is large, we can discard all but the first
term, $e_1$.  We can judge the accuracy of $\bar a$ by its variance
(assuming this is finite), which we can approximate as
\beq
  \Var_g(\bar a) & \approx & \Var_g(e_1) 
             \ \ =\ \ N^{-1} E_g\Big[\Big(w^{(i)}a(x^{(i)}\Big)^2\Big]
\label{eq-isvar0}
\eeq

We now return to an actual situation, in which $E_g[w^{(i)}]$ may not be one,
and $E_f[a]$ may not be zero, by modifying equation~(\ref{eq-isvar0}) 
suitably:
\beq
  \Var(\bar a) & \approx & N^{-1}\
    E_g\! \Big[\Big(w^{(i)}\, (a(x^{(i)})-E_f(a))\Big)^2 \Big]
       \ \Big/\ E_g\! \Big[\, w^{(i)} \,\Big]^2
\label{eq-isvar}\eeq
Geweke (1989) estimates this from the same data used to compute $\bar a$, 
as follows:
\beq
  \widehat{\Var}(\bar a) 
   & = & \sum_{i=1}^N \Big(w^{(i)}\, (a(x^{(i)})-\bar a)\Big)^2
         \ \Big/\ \Big[ \sum_{i=1}^N w^{(i)} \Big]^2
\label{eq-isvarest}\eeq
This is equivalent to the estimate discussed by Ripley (1987, Section~6.4) 
in the context of regenerative simulation.  When $N$ is small, Ripley
recommends using a jacknife estimate instead.

When $w^{(i)}$ and $a(x^{(i)})$ are independent under $g$,
equation~(\ref{eq-isvar}) simplifies to
\beq
  \Var_g(\bar a) 
  & \approx & N^{-1}\ E_g\Big[(w^{(i)})^2\Big]\,
                      E_g\Big[(a(x^{(i)})-E_f(a))^2\Big]
                      \ \Big/\ E_g\! \Big[\, w^{(i)} \,\Big]^2 \\[3pt]
  & = & N^{-1}\, \Big[\,1 + \Var_g\Big[w^{(i)}\,/\,E_g(w^{(i)})\Big]\,\Big]  
              \, \Var_f\Big[a(x^{(i)})\Big]
\label{eq-issamp}
\eeq
The last step above uses the following:
\beq
  \Var_f\Big[a(x^{(i)})\Big] \, =\, E_f\Big[(a(x^{(i)})-E_f(a))^2\Big] 
   & \!\!=\!\! & E_g\Big[w^{(i)}\,(a(x^{(i)}-E_f(a))^2\Big] 
                 \ \Big/\ E_g\Big[w^{(i)}\Big] \ \ \ \ \ \ \\[3pt]
   & \!\!=\!\! & E_g\Big[(a(x^{(i)})-E_f(a))^2\Big]
\eeq
Equation~(\ref{eq-issamp}) shows that when $w^{(i)}$ and $a(x^{(i)})$
are independent, the cost of using points drawn from $g(x)$ rather 
than $f(x)$ is given by one plus the variance of the normalized
importance weights.  We can estimate this using the sample variance
of $w_*^{(i)} = w^{(i)} \,/\, N^{-1} \sum w^{(i)}$.  This gives us
a rough indication of the factor by which the sample size is 
effectively reduced, without reference to any particular function
whose expectation is to be estimated.  Note that in many applications
the expectations of several functions will be estimated from the same 
sample of $x^{(i)}$.

The variance of the $w_*^{(i)}$ is also intuitively attractive as an
indicator of how accurate our estimates will be, since when it is
large, the few points with the largest importance weights will
dominate the estimates.  It would be imprudent to trust an estimate
when the adjusted sample size, $N\,/\,(1+\Var(w_*^{(i)}))$, is very
small, even if equation~(\ref{eq-isvarest}) gives a small estimate for
the variance of the estimator.  One should note, however, that it is
possible for the sample variance of the $w_*^{(i)}$ to be small even
when the estimates are wildly inaccurate, since this sample variance
could be a very bad estimate of the true variance of the normalized
importance weights.  This could happen, for example, if an important
mode of $f$ is almost never seen when sampling from $g$.

Earlier, it was suggested that $E_f[a]$ might be estimated by the
weighted average of the values of $a$ over the states of a Markov
chain that is started at each of the $x^{(i)}$.  The accuracy of such
an estimate should be estimated by treating these average values for
$a$ as single data points.  Treating the dependent states from along
the chain as if they were independently drawn from $g$ could lead to
overestimation of the effective sample size.

Finally, if the $x^{(i)}$ are not independently drawn from $g$, but
are instead generated by a Markov chain sampler, assessing the
accuracy of the estimates will be more difficult, as it will depend
both on the variance of the normalized importance weights and on the
autocorrelations produced by the Markov chain used.  This is one
reason for preferring a $p_n$ from which we can generate points
independently at the start of each annealed importance sampling run.

\section{Efficiency of annealed importance 
sampling}\vspace*{-10pt}\label{sec-eff}

The efficiency of annealed importance sampling depends on the
normalized importance weights, $w^{(i)}\,/\,E_g[w^{(i)}]$, not having
too large a variance.  There are several sources of variability in the
importance weights.  First, different annealing runs may end up in
different modes, which will be assigned different weights.  The
variation in weights due to this will be large if some important modes
are found only rarely.  There is no general guarantee that this will
not happen, and if it does, one can only hope to find a more effective
scheme for defining the annealing distributions, or use a radically
different Markov chain that eliminates the isolated modes altogether.

High variability in the importance weights can also result from using
transitions for each of these distributions that do not bring the
distribution close to equilibrium.  The extreme case of this is when
all the $T_j$ do nothing, in which case annealed importance sampling
reduces to simple importance sampling based on $p_n$, which will be
very inefficient if $p_n$ is not close to $p_0$.  Variability from
this source can reduced by increasing the number of iterations of the
basic Markov chain update used.  For example, if each $T_j$ consists
of $K$ Metropolis updates, the variance of the importance weights
might be reduced by increasing $K$, so that $T_j$ brings the state
closer to its equilibrium distribution, $p_j$ (at least within a local
mode).

Variability in the importance weights can also come from using a
finite number of distributions to interpolate between $p_0$ and $p_n$,
We can analyse how this affects the variance of the $w^{(i)}$ when the
sequence of distributions used comes from a smoothly-varying
one-parameter family, as in equation~(\ref{eq-family}).  For this
analysis, we will assume that each $T_j$ produces a state drawn from
$p_j$, independent of the previous state.  This assumption is of
course unrealistic, especially when there are isolated modes, but the
purpose here is to understand effects unrelated to Markov chain
convergence.

It is convenient to look at $\log(w^{(i)})$ rather than $w^{(i)}$ itself.
As discussed in Section~\ref{sec-acc}, we can measure the inefficiency
of estimation by one plus the variance of the normalized importance weights.
Using the fact that $E[Y^q]=\exp(q\mu+q^2\sigma^2/2)$ when $Y=\exp(X)$
and $X$ is Gaussian with mean $\mu$ and variance $\sigma^2$, we see
that if the $\log(w^{(i)})$ are Gaussian with mean $\mu$ and variance 
$\sigma^2$, the sample size will be effectively reduced by the factor
\beq
  1\,+\,\Var_g\!\left[{w^{(i)}\over E_g(w^{(i)})}\right] 
    \ =\ {E[(w^{(i)})^2] \over E[w^{(i)}]^2}
    \ =\ {\exp(2\mu+4\sigma^2/2) \over [\exp(\mu+\sigma^2/2)]^2}
    \ =\ \exp(\sigma^2)
\eeq
From equation~(\ref{eq-aisw}),
\beq
  \log(w^{(i)}) & = & 
  \sum_{j=1}^n\, \Big[ \log(f_{j-1}(x_{j-1})) \,-\, \log(f_j(x_{j-1})) \Big]
\eeq
If the distributions used are as defined by equation~(\ref{eq-family}),
\beq
  \log(w^{(i)}) & = & 
  \sum_{j=1}^n\ (\beta_{j-1}-\beta_j)\,
                \Big[ \log(f_0(x_{j-1})) - \log(f_n(x_{j-1})) \Big]
\eeq
If we further assume that the $\beta_j$ are equally spaced (between 0 and 1), 
we have
\beq
  \log(w^{(i)}) & = & {1 \over n}\
  \sum_{j=1}^n\ \Big[ \log(f_0(x_{j-1})) - \log(f_n(x_{j-1})) \Big]
\label{eq-aisvar}\eeq
Under the assumption that $T_j$ produces a state drawn independently
from $p_j$, and provided that $\log(f_0(x_{j-1})) - \log(f_n(x_{j-1}))$ has 
finite variance (when $x_{j-1}$ is drawn from $p_j$), the Central
Limit Theorem can be applied to conclude that $\log(w^{(i)})$ will 
have an approximately Gaussian distribution for large $n$ (keeping
$f_0$ and $f_n$ fixed as $n$ increases).  The variance of $\log(w^{(i)})$
will asymptotically have the form $\sigma_0^2/n$, for some constant 
$\sigma_0^2$, and one plus the variance of the normalized weights will 
have the form $\exp(\sigma_0^2/n)$. If we assume that each transition, 
$T_j$, takes a fixed amount of time (regardless of $n$), the time required 
to produce an estimate of a given degree of accuracy will be proportional to
$n\exp(\sigma_0^2/n)$, which is minimized when $n=\sigma_0^2$, at which
point the variance of the logs of the importance weights will be one and
the variance of the normalized importance weights will be $e\!-\!1$.

The same behaviour will occur when the $\beta_j$ are not equally
spaced, as long as they are chosen by a scheme that leads to
$\beta_{j-1} - \beta_j$ going down approximately in inverse proportion
to $n$.  Over a range of $\beta$ values for which $p_j$ is close to
Gaussian, and $p_0(x)$ is approximately constant in regions of high
density under $p_j$, an argument similar to that used for tempered
transitions (Neal 1996, Section 4.2) shows that the best scheme uses a
uniform spacing for $\log (\beta_j)$ (ie, a geometric spacing of the
$\beta_j$ themselves).  The results above also hold more generally for
annealing schemes that are based on families of distributions for
which the density at a given $x$ varies smoothly with a parameter
analogous to $\beta$.

We can get some idea of how the efficiency of annealed importance
sampling will be affected by the dimensionality of the problem by
supposing that under each $p_j$, the $K$ components of $x$ are
independent and identically distributed.  Assuming as above that each
$T_j$ produces an independent state drawn from $p_j$, the quantities
$\log(f_0(x_{j-1})) - \log(f_n(x_{j-1}))$ will be composed of $K$
identically distributed independent terms.  The variance of each such
quantity will increase in proportion to $K$, as will the variance of
$\log(w^{(i)})$, which will asymptotically have the form $K \sigma_0^2
/ n$.  The optimal choice of $n$ will be $K \sigma_0^2$, which makes
the variance of the normalized importance weights $e\!-\!1$, as above.
Assuming that behaviour is similar for more interesting distributions,
where the components are not independent, this analysis shows that
increasing the dimensionality of the problem will slow down annealed
importance sampling.  However, this linear slowdown is much less
severe than that for simple importance sampling, whose efficiency goes
down exponentially with $K$.

The above analysis assumes that each $T_j$ generates a state nearly
independent of the previous state, which would presumably require many
Metropolis or Gibbs sampling iterations.  It is probably better in
practice, however, to use transitions that do not come close to
producing an independent state, and hence take much less time, while
increasing the number of interpolating distributions to produce the
same total computation time.  The states generated would still come
from close to their equilibrium distributions, since these
distributions will change less from one annealing step to the next,
and the increased number of distributions may help to reduce the
variance of the importance weights, though perhaps not as much as in
the above analysis, since the terms in equation~(\ref{eq-aisvar}) will
no longer be independent.

We therefore see that the variance of the importance weights can be
reduced as needed by increasing the number of distributions used in
the annealing scheme, provided that the transitions for each
distribution are good enough at establishing equilibrium.  When there
are isolated modes, the latter provision will not be true in a global
sense, but transitions that sample well within a local mode can be
used.  Whether the performance of annealed importance sampling is
adequate will then depend on whether the annealing heuristic is in
fact capable of finding all the modes of the distribution.  In the
absence of any theoretical information pointing to where the modes are
located, reliance on some such heuristic is inevitable.

\section{Demonstrations on simple distributions}\vspace*{-10pt}\label{sec-demo}

To illustrate the behaviour of annealed importance sampling, I will
show how it works on a simple distribution with a single mode, using
Markov chain transitions that sample well for all intermediate
distributions, and on a distribution with two modes, which are
isolated with respect to the Markov chain transitions for the
distribution of interest.  Both distributions are over $R^6$.

In the unimodal distribution, the six components of the state, $x_1$
to $x_6$, are independent under $p_0$, with the distribution for each
being Gaussian with mean 1 and standard deviation 0.1.  This
distribution was defined by $f_0(x) = (1/2)
\sum_i\,(x_i\!-\!1)^2\,/\,0.1^2$, whose normalizing constant is
$(2\pi0.1^2)^{6/2}=0.000248$.  A sequence of annealing distributions
was defined according to the scheme of equation~(\ref{eq-family}).
Under the distribution chosen for $p_n$, the components were
independent, each being Gaussian with mean zero and standard deviation
1.  The function $f_n$ used to define this distribution was chosen to
be the corresponding Gaussian probability density, which was
normalized.  We can therefore estimate the normalizing constant for
$f_0$ by the average of the importance weights.

To use annealed importance sampling, we must choose a sequence of
$\beta_j$ that define the intermediate distributions.  Both the number
and the spacing of the $\beta_j$ must be appropriate for the problem.
As mentioned in the previous section, for a Gaussian $p_0$, and a
diffuse $p_n$, we expect that a geometric spacing will be appropriate
for the $\beta_j$ that are not too far from one.  I spaced the
$\beta_j$ near zero arithmetically.  In detail, for the first test, I
used 40 $\beta_j$ spaced uniformly from 0 to 0.01, followed by 160
$\beta_j$ spaced geometrically from 0.01 to 1, for a total of 200
distributions.  In later tests, annealing sequences with twice as many
and half as many distributions were also used, spaced according to the
same scheme.

We must also define Markov chain transitions, $T_j$, for each of these
distributions.  In general, one might use different schemes for
different distributions, but in these tests, I used Metropolis updates
with the same proposal distributions for all $T_j$ (the transition
probabilities themselves were of course different for each $T_j$,
since the Metropolis acceptance criterion changes).  In detail, I used
sequences of three Metropolis updates, with Gaussian proposal
distributions centred on the current state having covariances of
$0.05^2I$, $0.15^2I$, and $0.5^2I$.  Used together, these three
proposal distributions lead to adequate mixing for all of the
intermediate distributions.  For the first test, this sequence of
three updates was repeated 10 times to give each $T_j$; in one later
test, it was repeated only 5 times.

For each test, 1000 annealing runs were done.  In the first test, 200
states were produced in each run, as a result of applying each $T_j$
in succession, starting from a point generated independently from
$p_{200}$.  I saved only every twentieth state, however, after
applying $T_{180}$, $T_{160}$, etc.\ down to $T_{0}$.  Note that $T_0$
was applied at the end of each run in these tests, even though this is
not required (this occurs naturally with the program used).  Only the
state after applying $T_0$ was used for the estimates, even though it
is valid to use the state after $T_1$ as well.

Figure~\ref{fig-uni1} shows the results of this first test.  The upper
graphs show how the variance of the log of the importance weights
increases during the course of a run.  (Importance weights before the
run is over are defined as in equation~(\ref{eq-aisw}), but with the
factors for the later distributions omitted.)  When, as here, the
transitions for all distributions are expected to mix well, the best
strategy for minimizing the variance of the final weights is to space
the $\beta_j$ so that the variance of the log weights increases by an
equal amount in each annealing step.  The plot in the upper right
shows that the spacing chosen for this test is close to optimal in
this respect.  Furthermore, according to the analysis of
Section~\ref{sec-eff}, the number of intermediate distributions used
here is close to optimal, since the variance of the logs of the
weights at the end of the annealing run is close to one.

The lower two graphs in Figure~\ref{fig-uni1} show the distribution of
the value of the first component of the state ($x_1$) in this test.
As seen in the lower left, this distribution narrows to the
distribution under $p_0$ as $\beta$ approaches one.  The plot in the
lower right shows the values of the first component and of the
importance weights for the states at the ends of the runs.  In this
case, the values and the weights appear to be independent.

The estimate for the expectation of the first component of the state
in this first test is 1.0064, with standard error 0.0050, as estimated
using equation~(\ref{eq-isvarest}).  This is compatible with the true
value of one.  In this case, the error estimate from
equation~(\ref{eq-isvarest}) is close what one would arrive at from
the estimated standard deviation of 0.10038 and the adjusted sample
size of $N\,/\,(1\!+\!\Var(w_*)) = 1000\,/\,(1\!+\!1.12) = 472$, as
expected when the values and the weights are independent.  The average
of the importance weights for this test was 0.000236, with standard
error 0.000008 (estimated simply from the sample variance of the
weights divided by $N$); this is compatible with the true normalizing
constant of 0.000248.

Two tests were done in which each run used half as much computer time
as in the first test.  In one of these, the annealing sequence was
identical to the first test, but the number of repetitions of the
three Metropolis updates in each $T_j$ was reduced from 10 to 5.  This
increased the variance of the normalized importance weights to 2.18,
with a corresponding increase in the standard errors of the estimates.
In the other test, the number of distributions in the annealing
sequence was cut in half (spaced according to the same scheme as
before), while the number of Metropolis repetitions was kept at 10.
This increased the variance of the normalized importance weights to
2.72.  As expected, spreading a given number of updates over many
intermediate distributions appears to be better than using many
updates to try to produce nearly independent points at each of fewer
stages.

The final test on this unimodal distribution used twice as many
intermediate distributions, spaced according to the same scheme as
before.  This reduced the variance of the normalized importance
weights to 0.461, with a corresponding reduction in standard errors,
but the benefit in this case was not worth the factor of two increase
in computer time.  However, this test does confirm that when each
$T_j$ mixes well, the variance of the importance weights can be
reduced as desired by spacing the $\beta_j$ more closely.

Tests were also done on a distribution with two modes, which was a
mixture of two Gaussians, under each of which the six components were
independent, with the same means and standard deviations.  One of
these Gaussians, with mixing proportion $1/3$, had means of 1 and
standard deviations of 0.1, the same as the distribution used in the
unimodal tests.  The other Gaussian, with mixing proportion $2/3$, had
means of $-1$ and standard deviations of 0.05.  This mixture
distribution was defined by the following $f_0$: 
\beq
  f_0(x) & = &
      \exp\left[-{1\over2}\sum_{i=1}^6 {(x_i-1)^2\over0.1^2}\right]
      \ +\ 128\,\exp\left[-{1\over2}\sum_{i=1}^6 {(x_i+1)^2\over0.05^2}\right]
\eeq
The normalizing constant for this $f_0$ is $3\,(2\pi0.1^2)^{6/2}=0.000744$.
The means of the components with respect to this $p_0$ are $-1/3$.

The same $f_n$ as before was used for these tests (independent
standard Gaussian distributions for each component, normalized).  The
same transitions based on Metropolis updates were used as well, along
with the same scheme for spacing the $\beta_j$.  For the first test,
the number of distributions used was 200, as in the first test on the
unimodal distribution.

The results are shown in Figure~\ref{fig-multi1}.  As seen in the
lower left of the figure, the distributions for $\beta$ near zero
cover both modes, but as $\beta$ is increased, the two modes become
separated.  The Metropolis updates are not able to move between these
modes when $\beta$ is near one, even when using the larger proposals
with standard deviation 0.5, since the probability of proposing a
movement to the other mode simultaneously for all six components is
very small.  Both modes are seen when annealing, but the mode at $-1$
is seen only rarely --- 27 times in the 1000 runs --- despite the fact
that it has twice the probability of the other mode under the final
distribution at $\beta=1$.  An unweighted average over the final
states of the annealing runs would therefore give very inaccurate
results.

The plot in the lower right of the figure shows how the importance
weights compensate for this unrepresentative sampling.  The runs that
ended in the rarely-sampled mode received much higher weights than
those ending in the well-sampled mode.  The estimate for the
expectation of the first component from these runs was $-0.363$, with an
estimated standard error of 0.107 (from equation~(\ref{eq-isvarest})),
which is compatible with the true value of $-1/3$.  This standard
error estimate is less than one might expect from the estimated
standard deviation of 0.92 and the adjusted sample size of
$N\,/\,(1\!+\!\Var(w_*))$, which was 35.0.  The difference arises
because the values and the importance weights are not independent in
this case.

\input{fig-uni1.tex}
\input{fig-multi1.tex}

The average of the importance weights in these runs was 0.000766, with
an estimated standard error of 0.000127, which is compatible with the
true value of 0.000744 for the normalizing constant of $f_0$.

We therefore see that annealed importance sampling produces valid
estimates for this example.  However, the procedure is less efficient
than we might hope, because so few runs end in the mode at $-1$.
Another symptom of the problem is that the variance of the normalized
importance weights in this test was 27.6 --- quite high compared to
the variance of 1.12 seen in the similar test on the unimodal
distribution.  We can see how this comes about from the upper plots in
Figure~\ref{fig-multi1}.  For small values of $\beta$, these plots are
quite similar to those in Figure~\ref{fig-uni1}, presumably because
the mode at $-1$ has almost no influence for these distributions.
However, this mode becomes important as $\beta$ approaches one,
producing a high variance for the weights at the end.

One might hope to reduce the variance of the importance weights by
increasing the number of intermediate distributions (ie, by spacing
the $\beta_j$ more closely).  I ran tests with twice as many
distributions, and with four times as many distributions, in both
cases using the same number of Metropolis updates for each
distribution as before.  The results differed little from those in the
first test.  The variance of the importance weights for runs ending
within each mode was reduced, but the difference in importance weights
between modes was not reduced, and the number of runs ending in the
mode at $-1$ did not increase.  There was therefore little difference
in the standard errors for the estimates.

For this example, the annealing heuristic used was only marginally
adequate.  One could expect to obtain better results only by finding a
better initial distribution, $p_n$, or a better scheme for
interpolating from $p_n$ to $p_0$ than that of
equation~(\ref{eq-family}).  This example also illustrates the dangers
of uncritical reliance on empirical estimates of accuracy.  If only
100 runs had been done, the probability that \emph{none} of the runs
would have found the mode at $-1$ would have been around $0.07$.  This
result can be simulated using the first 100 runs that ended in the
mode at $+1$ from the 1000 runs of the actual test.  Based on these
100 runs, the estimate for the expectation of the first component is
0.992, with an estimated standard error 0.017, and the estimate for
the normalizing constant of $f_0$ is 0.000228, with an estimated
standard error of 0.000020.  Both estimates differ from the true
values by many times the estimated standard error.  Such unrecognized
inaccuracies are of course also possible with any other importance
sampling or Markov chain method, whenever theoretically-derived
guarantees of accuracy are not available.

\section{Demonstration on a linear regression problem
}\vspace*{-10pt}\label{sec-lin}

To illustrate the use of annealed importance sampling for statistical
problems, I will briefly describe its application to two Bayesian
models for a linear regression problem, based on Gaussian and Cauchy
priors.  This example, and that of the previous section, are
implemented using my software for flexible Bayesian modeling (version
of \mbox{1998-09-01}).  The data and command files used are included
with that software, which is available from my web page.

The data consists of 100 independent cases, each having 10
real-valued predictor variables, $x_1,\ldots,x_{10}$ and a real-valued
response variable, $y$, which is modeled by 
\begin{eqnarray*}
  y & = & \sum_{k=1}^{10} \beta_k\, x_i \ +\ \epsilon
\end{eqnarray*}
The residual, $\epsilon$, is modeled as Gaussian with mean zero
and unknown variance $\sigma^2$.  The 100 cases were synthetically generated 
from this model with $\sigma^2=1$ and with $\beta_1=1$, $\beta_2=0.5$, 
$\beta_3=-0.5$, and $\beta_k=0$ for $4 \le i \le 10$.  The predictor variables 
were generated from a multivariate Gaussian with the variance of each $x_i$ 
being one and with correlations of 0.9 between each pair of $x_i$.

Two Bayesian models were tried.  In both, the prior for the reciprocal
of the residual variance ($1/\sigma^2$) was gamma with mean $1/0.1^2$
and shape parameter $0.5$.  Both models also had a hyperparameter,
$\nu^2$, controlling the width of the distribution of the $\beta_k$.
Its reciprocal was given a gamma prior with mean $1/0.05^2$ and shape
parameter $0.25$.  For the model with Gaussian priors, $\nu^2$ was the
variance of the $\beta_k$, which had mean zero, and were independent
conditional on $\nu^2$.  The model based on Cauchy priors was similar,
except that $\nu$ was the width parameter of the Cauchy distribution
(ie, the density for $\beta_k$ conditional on $\nu$ was $(1/\pi\nu) [1
+ \beta_k^2/\nu^2]^{-1}$).  One might suspect that the Cauchy prior will
prove more appropriate for the actual data, since this prior gives substantial
probability to situations where many of the $\beta_k$ are close to
zero, but a few $\beta_k$ are much bigger.

It seems quite possible that the posterior using the Cauchy prior
could be multimodal.  Since the $x_i$ are highly correlated, one
$\beta_k$ can to some extent substitute for another.  The Cauchy prior
favours situations where only a few $\beta_k$ are large.  This could
produce several posterior modes that correspond to different sets of
$\beta_k$ being regarded as significant. 

I sampled for both models using a combination of Gibbs sampling for
$\sigma^2$ and the ``hybrid Monte Carlo'' method for the $\beta_k$
(see Neal 1996).  There was no sign of any problems with isolated
modes, but it is difficult to be sure on this basis that no such modes
exist.  Annealed importance sampling was applied in order to either
find any isolated modes or provide further evidence of their absence,
and also to compare the two models by calculating their marginal
likelihoods.

An annealing schedule based on equation~(\ref{eq-family}) was used.
After some experimentation, adequate results were obtained using such
a schedule with 1000 distributions:\ \ 50 distributions geometrically
spaced from $\beta=10^{-8}$ to $\beta=10^{-6}$, then 450 distributions
geometrically spaced from $\beta=10^{-6}$ to $\beta=0.05$, and finally
500 distributions geometrically spaced from $\beta=0.05$ to $\beta=1$.
Hybrid Monte Carlo updates were used for each distribution.  A single
annealing run took approximately 8 seconds on our 194~MHz SGI
machine.  I did 500 such runs for each model.

Because a few of the annealing runs resulted in much smaller weights
than others, the variance of the logs of the weights was very large,
and hence was not useful in judging whether the annealing schedule was
good.  Instead, I looked at $W = \log(1+\mbox{Var}(w^{(i)}_*))$, the log
of one plus the variance of the normalized importance weights.  If the
distribution of the logs of the weights were Gaussian, $W$ would be
equal to the variance of the logs of the weights.  When this
distribution is not Gaussian, $W$ is less affected by a few extremely
small weights.  Plots of $W$ show that for both models it increases
approximately linearly with the index of the distribution, reaching a
final value around 0.6, only a bit less than the optimal value of one.

For both models, the estimates of the posterior means of the $\beta_k$
found using annealed importance did not differ significantly from those
found using hybrid Monte Carlo without annealing.  It therefore
appears that isolated modes were not present in this problem.  The
annealed importance sampling runs yielded estimates for the log of the
marginal likelihood for the model with Gaussian priors of -158.68 and
for the model with Cauchy priors of -158.24, with a standard error of
0.04 for both estimates.  The difference of 0.44 corresponds to a
Bayes factor of 1.55 in favour of the model with Cauchy priors.

\section{Relationship to tempered transitions}\vspace*{-10pt}\label{sec-tt}

Several ways of modifying the simulated annealing procedure in order
to produce asymptotically correct estimates have been developed in the
past, including simulated tempering (Marinari and Parisi 1992; Geyer
and Thompson 1995) and Metropolis coupled Markov chains (Geyer 1991).
The method of tempered transitions (Neal 1996) is closely related to
the annealed importance sampling method of this paper.

The tempered transition method samples from a distribution of
interest, $p_0$, using a Markov chain whose transitions are defined in
terms of an elaborate proposal procedure, involving a sequence of
other distributions, $p_1$ to $p_n$.  The proposed state is found by
simulating a sequence of base transitions, $\Tu_1$ to $\Tu_n$, which
leave invariant the distributions $p_1$ to $p_n$, followed by a second
sequence of base transitions, $\Td_n$ to $\Td_1$, which leave $p_n$ to
$p_1$ invariant, and which are the reversals of the corresponding
$\Tu_j$ with respect to the $p_j$.  The decision whether to accept or
reject the final state is based on a product of ratios of
probabilities under the various distributions; if the proposed state is
rejected, the new state is the same as the old state.

In detail, such a tempered transition operates as follows, starting from
state $\xu_0$:
\beq\begin{array}{l}
  \mbox{Generate $\xu_1$ from $\xu_0$ using $\Tu_1$.} \\[3pt]
  \mbox{Generate $\xu_2$ from $\xu_1$ using $\Tu_2$.} \\[3pt]
  \mbox{\hspace*{70pt}\ldots} \\[3pt]
  \mbox{Generate $\xb_n$ from $\xu_{n-1}$ using $\Tu_n$.} \\[3pt]
  \mbox{Generate $\xd_{n-1}$ from $\xb_n$ using $\Td_n$.} \\[3pt]
  \mbox{\hspace*{70pt}\ldots} \\[3pt]
  \mbox{Generate $\xd_1$ from $\xd_2$ using $\Td_2$.} \\[3pt]
  \mbox{Generate $\xd_0$ from $\xd_1$ using $\Td_1$.}
\end{array}\label{eq-tt}\eeq
The state $\xd_0$ is then accepted as the next state of the Markov chain
with probability
\beq
 \min\!  \left[1,\
       {p_1(\xu_0) \over p_0(\xu_0)} \cdots 
       {p_n(\xu_{n-1}) \over p_{n-1}(\xu_{n-1})} \!\cdot\!
       {p_{n-1}(\xd_{n-1}) \over p_n(\xd_{n-1})} \cdots 
       {p_0(\xd_0) \over p_1(\xd_0)} 
       \right]
\label{eq-tta}\eeq

The second half of the tempered transition procedure~(\ref{eq-tt}) is
identical to the annealed importance sampling
procedure~(\ref{eq-samp}), provided that $\Td_n$ in fact generates a
point from $p_n$ that is independent of $\xb_n$.  We can also
recognize that the annealed importance sampling weight given by
equation~(\ref{eq-aisw}) is essentially the same as the second half of
the product defining the tempered transition acceptance
probability~(\ref{eq-tta}).  Due to these similarities, the
characteristics of annealed importance sampling will be quite similar
to those of the corresponding tempered transitions.  In particular,
the comparison by Neal (1996) of tempered transitions with simulated
tempering is relevant to annealed importance sampling as well.

The major difference between annealed importance sampling and tempered
transitions is that each tempered transition requires twice as much
computation as the corresponding annealing run, since a tempered
transition involves an ``upward'' sequence of transitions, from $p_1$
to $p_n$, as well as the ``downward'' sequence, from $p_n$ to $p_1$,
that is present in both methods.  This is a reason to prefer annealed
importance sampling when it is easy to generate independent points
from the distribution $p_n$.  When this is not easy, tempered
transitions might be preferred, though annealed importance sampling
could still be used in conjunction with a Markov chain sampler that
produces dependent points from $p_n$.  With tempered transitions,
there is also the possibility of using more than one sequence of
annealing distributions (with the sequence chosen randomly for each
tempered transition, or in some fixed order).  Potentially, this could
lead to good sampling even when neither annealing sequence would be
adequate by itself.  There appears to be no way of employing multiple
annealing sequences with annealed importance sampling without adding
an equivalent of the ``upward'' sequence present in tempered
transitions.

When tempered transitions are used, the idea behind annealed
importance sampling can be applied in order to estimate ratios of
normalizing constants, which were previously unavailable when using
tempered transitions.  To see how to do this, note that the first half
of a tempered transition (up to the generation of $\xu_{n-1}$ from
$\xu_{n-2}$ using $\Tu_{n-1}$) is the same as an annealed importance
sampling run, but with the sequence of distributions reversed ($p_0$
and $p_n$ exchange roles, the first state of the run is the current
state, $\xu_0$, which comes from $p_0$, and in general, $x_j$
of~(\ref{eq-samp}) corresponds to $\xu_{n-1-j}$ of~(\ref{eq-tt})).  The
importance weights for this backwards annealed importance sampling are
\beq
  \hat w^{(i)} & = & {f_1(\xu_0) \over f_0(\xu_0)}\,
                {f_2(\xu_1) \over f_1(\xu_1)}\,
                \cdots\,  
                {f_{n-1}(\xu_{n-2}) \over f_{n-2}(\xu_{n-2})}\,
                {f_n(\xu_{n-1}) \over f_{n-1}(\xu_{n-1})}
\label{eq-normu}\eeq
The average of these weights for all tempered transitions (both
accepted and rejected) will converge to 
$\int\!f_n(x)\,dx\,/\,\int\!f_0(x)\,dx$, the ratio of normalizing
constants for $f_n$ and $f_0$.  

A similar estimate can be found by imagining the reversal of the
Markov chain defined by the tempered transitions. In this chain, the
states are visited in the reverse order, the accepted transitions of
the original chain become accepted transitions in the reversed chain
(but with the reversed sequence of states), and the rejected
transitions of the original chain remain unchanged.  An importance
sampling estimate for the ratio of normalizing constants for $f_n$
and $f_0$ can be obtained using this reversed chain, in the same
manner as above.  The importance weights for the accepted transitions 
are as follows, in terms of the original chain:
\beq
  \check w^{(i)} & = & {f_1(\xd_0) \over f_0(\xd_0)}\,
                {f_2(\xd_1) \over f_1(\xd_1)}\,
                \cdots\,  
                {f_{n-1}(\xd_{n-2}) \over f_{n-2}(\xd_{n-2})}\,
                {f_n(\xd_{n-1}) \over f_{n-1}(\xd_{n-1})}
\label{eq-normd}\eeq
The importance weights for the rejected transitions are the same as
in equation~(\ref{eq-normu}). These two estimates can be averaged, 
producing an estimate that uses the states at both the beginning and
the end of the accepted transitions, plus the states at the beginning 
of the rejected transitions, with double weight.

An estimate for the ratio of the normalizing constant for $f_j$ to
that for $f_0$ can be found in similar fashion for any of the
intermediate distributions, by simply averaging the weights obtained
by truncating the products in equations~(\ref{eq-normu})
and~(\ref{eq-normd}) at the appropriate point.  These weights can also
be used to estimate expectations of functions with respect to these
intermediate distributions.  Note that error assessment for all these
importance sampling estimates will have to take into account both the
variance of the importance weights and the autocorrelations produced
by the Markov chain based on the tempered transitions.

A cautionary note regarding these estimates comes from considering the
situation when only two distributions are used, which are the prior
and the posterior for a Bayesian model.  The estimate for the
reciprocal of the marginal likelihood based on
equation~(\ref{eq-normu}) will then be the average over points drawn
from the posterior of the reciprocal of the likelihood.  This
estimator will often have infinite variance, and will be very bad for
any problem where there is enough data that the posterior is not much
affected by the prior (since the marginal likelihood \emph{is}
affected by the prior).  Compare this to the annealed importance
sampling estimate for the marginal likelihood using just these two
distributions, which will be the average of the likelihood over points
drawn from the prior.  This is not very good when the posterior is
much more concentrated than the prior, but it is not as bad as
averaging the reciprocal of the likelihood.  Even when many
intermediate distributions are used, it seems possible the annealed
importance sampling estimates may be better than the corresponding
``backwards'' estimates using tempered transitions (assuming that
$p_n$ is more diffuse than $p_0$).

\section{Relationship to sequential importance 
         sampling}\vspace*{-10pt}\label{sec-sis}

A variant of sequential importance sampling recently developed by
MacEachern, Clyde, and Liu (1998) can be viewed as an instance of
annealed importance sampling, in which the sequence of distributions
is obtained by looking at successively more data points.

This method (which MacEachern, \textit{et al} call Sequential
Importance Sampler S4) applies to a model for the joint distribution
of observable variables $x_1,\ldots,x_n$ along with associated latent
variables $s_1,\ldots,s_n$ (which have a finite range).  We are able
to compute these joint probabilities, as well as the marginal
probabilities for the $x_k$ together with the $s_k$ over any subset of
the indexes.  We wish to estimate expectations with respect to the
conditional distribution of $s_1,\ldots,s_n$ given known values for
$x_1,\ldots,x_n$.  We could apply Gibbs sampling to this problem, but
it is possible that it will be slow to converge, due to isolated
modes.

The method of MacEachern, \textit{et al} can be viewed as annealed
importance sampling with a sequence of distributions, $p_0$ to $p_n$,
in which $p_j$ is related to the distribution conditional on $n\!-\!j$ 
of the observed variables; $p_0$ is then the distribution of interest,
conditional on all of $x_1,\ldots,x_n$.  In detail, these distributions 
have probabilities proportional to the following $f_j$:
\beq
  \lefteqn{f_j(s_1,\ldots,s_n)}\ \ \ \ \nonumber\\
  & = & P(s_1,\ldots,s_{n-j},\,x_1,\ldots,x_{n-j})
  \prod_{\!\!\!k=n-j+1\!\!\!}^n \!
      P(s_k\ |\ x_1,\ldots,x_k,\,s_1,\ldots,s_{k-1})\ \ \ \ \ \ \ \ 
\label{eq-sisanneal}\eeq

We can apply annealed importance sampling with this sequence of
distributions, using transitions defined as follows. $T_j$ begins with
some number of Gibbs sampling updates for $s_1$ to $s_{n-j}$, based
only on $P(s_1,\ldots,s_{n-j}\ |\ x_1,\ldots,x_{n-j})$.  We can ignore
$s_{n-j+1}$ to $s_n$ here because we can generate values for them
afterward from their conditional distribution (under $f_j$) given
$s_1$ to $s_{n-j}$, independently of their previous values.  This is
done by forward simulation based on their conditional probabilities.
(Actually, there is no need to generate values for $s_k$ with
$k>n\!-\!j\!+\!1$, since these values have no effect on the subsequent
computations anyway.)  This is easily seen to be equivalent to the
sampling done in procedure S4 of MacEachern, \mbox{\textit{et al}}.

The importance weights of equation~(\ref{eq-aisw}) are products of 
factors of the following form: 
\beq 
  \lefteqn{{f_{j-1}(s_1,\ldots,s_n) \over f_j(s_1,\ldots,s_n)}}
  \ \ \ \ \ \ \ \nonumber\\
  & \!=\! &
 { P(s_1,\ldots,s_{n-j+1},\,x_1,\ldots,x_{n-j+1}) \over
   P(s_1,\ldots,s_{n-j},\,x_1,\ldots,x_{n-j})\
   P(s_{n-j+1}\ |\ x_1,\ldots,x_{n-j+1},\,s_1,\ldots,s_{n-j}) }
  \ \ \ \ \ \ \ \ \\[4pt]
  & \!=\! &
  { P(s_{n-j+1},\,x_{n-j+1}\ |\ x_1,\ldots,x_{n-j},\,s_1,\ldots,s_{n-j})
    \over
    P(s_{n-j+1}\ |\ x_1,\ldots,x_{n-j+1},\,s_1,\ldots,s_{n-j}) } \\[4pt]
  & \!=\! &
    P(x_{n-j+1}\ |\ x_1,\ldots,x_{n-j},\,s_1,\ldots,s_{n-j})
\eeq
The product of these factors produces the same weights as used 
by MacEachern, \mbox{\textit{et al}}.

Sequential Importance Sampler S4 of MacEachern, \textit{et al} is thus
equivalent to annealed importance sampling with the annealing
distributions defined by equation~(\ref{eq-sisanneal}).  Unlike the
family of distributions given by equation~(\ref{eq-family}), these
distributions form a fixed, discrete family.  Consequently, the
variance of the importance weights cannot be decreased by increasing
the number of distributions.  This could sometimes make the method too
inefficient for practical use.  However, it is possible that the
sequence of distributions defined by equation~(\ref{eq-sisanneal})
could be extended to a continuous family by partially conditioning on
the $x_k$ in some way (eg, by adjusting the variance in a Gaussian
likelihood).  Other forms of annealed importance sampling (eg, based
on the family of equation~(\ref{eq-family})) could also be applied to
this problem.

\section{Discussion}\vspace*{-10pt}\label{sec-disc}

Annealed importance sampling is potentially useful as a way of dealing
with isolated modes, as a means of calculating ratios of normalizing
constants, and as a general Monte Carlo method that combines
independent sampling with the adaptivity of Markov chain methods.  

Handling isolated modes was the original motivation for annealing, and
has been the primary motivation for developing methods related to
annealing that produce asymptotically correct results.  Annealed
importance sampling is another such method, whose characteristics are
similar to those of tempered transitions.  As I have discussed (Neal
1996), which of these methods is best may depend on whether the
sequence of annealing distributions is ``deceptive'' in certain ways.
It is therefore not possible to say that annealed importance sampling
will always be better than other methods such as simulated tempering,
but it is probably the most easily implemented of these methods.

Annealing methods are closely related to methods for estimating ratios
of normalizing constants based on simulations from many distributions,
many of which are discussed by Gelman and Meng (1998).  It is
therefore not surprising that the methods of simulated tempering
(Marinari and Parisi 1992; Geyer and Thompson 1995) and Metropolis
coupled Markov chains (Geyer 1991) easily yield estimates for ratios
of normalizing constants as a byproduct.  Tempered transitions were
previously seen as being deficient in this respect (Neal 1996), but we
now see that such estimates can in fact be obtained by using annealed
importance sampling estimators in conjunction with tempered
transitions.  One can also estimate expectations with respect
to all the intermediate distributions in this way (as is also possible
with simulated tempering and Metropolis coupled Markov chains).

Ratios of normalizing constants can also be obtained when using
annealed importance sampling itself, which from this perspective can
be seen as a form of thermodynamic integration (see Gelman and Meng
1998).  One might expect a thermodynamic integration estimate based on
a finite number of points to suffer from systematic error, but the
results of this paper show that the annealed importance sampling
estimate for the ratio of normalizing constants is in fact unbiased,
and will converge to the correct value as the number of annealing runs
increases.  (Note that in this procedure one averages the estimates
from multiple runs for the ratio of normalizing constants, not for the
log of this ratio, as might perhaps seem more natural.)

Unlike simulated tempering and the related method of umbrella sampling
(Torrie and Valleau 1977), no preliminary estimates for ratios of
normalizing constants are required when using annealed importance
sampling.  Metropolis coupled Markov chains share this advantage, but
have the disadvantage that they require storage for states from all
the intermediate distributions.  Annealed importance sampling may
therefore be the most convenient general method for estimating
normalizing constants.

In addition to these particular uses, annealed importance sampling may
sometimes be attractive because it combines independent sampling with
the ability of a Markov chain sampler to adapt to the characteristics
of the distribution.  Evans (1991) has also devised an adaptive
importance sampling method that makes use of a sequence of
intermediate distributions, similar to that used for annealing.  His
method requires that a class of tractable importance sampling
densities be defined that contains a density appropriate for each of
the distributions in this sequence.  Annealed importance sampling
instead uses a sampling distribution that is implicitly defined by the
operation of the Markov chain transitions, whose density is generally
not tractable to compute, making its use for simple importance
sampling infeasible.  From this perspective, the idea behind annealed
importance sampling is that one can nevertheless find appropriate
importance weights for use with this sampling distribution by looking
at ratios of densities along the sequence of intermediate
distributions.

One annoyance with Markov chain Monte Carlo is the need to estimate
autocorrelations in order to assess the accuracy of the estimates
obtained.  Provided the points from $p_n$ used to start the annealing
runs are generated independently, there is no need to do this with
annealed importance sampling.  Instead, one must estimate the variance
of the normalized importance weights.  This may perhaps be easier,
though nightmare scenarios in which drastically wrong results are
obtained without there there being any indication of a problem are
possible when using methods of either sort.  For annealed importance
sampling, this can occur when the distribution of the importance
weights has a heavy upward tail that is not apparent from the data
collected.

Another annoyance with Markov chain Monte Carlo is the need to decide
how much of a run to discard as ``burn-in'' --- ie, as not coming from
close to the equilibrium distribution.  If only one, long run is
simulated, the exact amount discarded as burn-in may not be crucial,
but if several shorter runs are done instead, as is desirable in order
to diagnose possible non-convergence, the decision may be harder.
Discarding too little will lead to biased estimates; discarding too
much will waste data.  With annealed importance sampling, one must
make an analogous decision of how much computation time to spend on
the annealing runs themselves, which determine the importance weights,
and how much to spend on simulating a chain that samples from $p_0$
starting from the final state from the annealing run (as is usually
desirable, see Section~\ref{sec-ais}).  However, this decision affects
only the variance of the estimates --- the results are asymptotically
correct regardless of how far the annealing process is from reaching
equilibrium.

Regenerative methods (Mykland, Tierney, and Yu 1995) also eliminate
the problems of dealing with sequential dependence (and also replace
them with possible problems due to heavy-tailed distributions).  To
use regenerative methods, an appropriate ``splitting'' scheme must be
devised for the Markov chain sampler.  For high-dimensional problems,
this may be harder than defining an appropriate sequence of
intermediate distributions for use with annealed importance sampling.

As discussed in Section~\ref{sec-eff}, the time required for annealed
importance sampling can be expected to increase in direct proportion
to the dimensionality of the problem (in addition to any increase due
to the Markov chain samplers used being slower in higher dimensions).
One must also consider the human and computer time required to select
an appropriate sequence of intermediate distributions, along with
appropriate Markov chain transitions for each.  For these reasons,
annealed importance sampling will probably be most useful when it
allows one to find needed ratios of normalizing constants, or serves
to avoid problems with isolated modes.  One should note, however, that
the potential for problems with multiple modes exists whenever there
is no theoretical guarantee that the distribution is unimodal.

\section*{Acknowledgements}\vspace*{-10pt}

I thank David MacKay for helpful comments. This research was supported
by the Natural Sciences and Engineering Research Council of Canada.

\section*{References}\vspace*{-10pt}

\leftmargini 0.2in
\labelsep 0in

\begin{description}
\itemsep 2pt

\item
  Evans, M.\ (1991) ``Chaining via annealing'', {\em Annals of Statistics},
  vol.~19, pp.~382-393.

\item
  Gelman, A.\ and Meng, X.-L.\ (1998) ``Simulating normalizing constants:
  From importance sampling to bridge sampling to path sampling'', 
  \textit{Statistical Science}, vol.~13, pp.~163-185.

\item
  Geyer, C.~J.\ (1991) ``Markov chain Monte Carlo maximum likelihood'',
  in E.~M.~Keramidas (editor), {\em Computing Science and Statistics:
  Proceedings of the 23rd Symposium on the Interface}, pp.~156-163,
  Interface Foundation.

\item
  Geyer, C.~J.\ and Thompson, E.~A.\ (1995) ``Annealing Markov chain Monte
  Carlo with applications to ancestral inference'', {\em Journal of the
  American Statistical Association}, vol.~90, pp.~909-920.

\item
  Geweke, J.\ (1989) ``Bayesian inference in econometric models using
  Monte Carlo integration'', {\em Econometrica}, vol.~57, pp.~1317-1339.

\item
  Gilks, W.~R., Richardson, S., and Spiegelhalter, D.~J.\ (1996)
  {\em Markov Chain Monte Carlo in Practice}, London: Chapman and Hall.

\item
  Hastings, W.~K.\ (1970) ``Monte Carlo sampling methods using Markov chains 
  and their applications'', {\em Biometrika}, vol.~57, pp.~97-109.

\item
  Jarzynski, C.\ (1997a) ``Nonequilibrium equality for free energy 
  differences'', {\em Physical Review Letters}, vol.~78, pp.~2690-2693.

\item
  Jarzynski, C.\ (1997b) ``Equilibrium free-energy differences from
  nonequilibrium measurements: A master-equation approach'', {\em Physical
  Review E}, vol.~56, pp.~5018-5035.

\item
  Kirkpatrick, S., Gelatt, C.~D., and Vecchi, M.~P.\ (1983) ``Optimization
  by simulated annealing'', {\em Science}, vol.~220, pp.~671-680.

\item
  Marinari, E.\ and Parisi, G.\ (1992) ``Simulated tempering: A new
  Monte Carlo scheme'', {\em Europhysics Letters}, vol.~19, pp.~451-458.

\item
  MacEachern, S.~N., Clyde, M., and Liu, J.~S. (1998) ``Sequential
  importance sampling for nonparametric Bayes models:\ The next generation'',
  to appear in {\em The Canadian Journal of Statistics}.

\item
  Metropolis, N., Rosenbluth, A.~W., Rosenbluth, M.~N., Teller, A.~H., 
  and Teller, E.\ (1953) ``Equation of state calculations by fast computing 
  machines'', {\em Journal of Chemical Physics}, vol.~21, pp.~1087-1092.

\item
  Mykland, C., Tierney, L., and Yu, B.\ (1995). ``Regeneration in Markov Chain
  Samplers'', {\em Journal of the American Statistical Association}, 
  vol.~90, pp.~233-241. 

\item
  Neal, R.~M.\ (1996) ``Sampling from multimodal distributions using
  tempered transitions'', {\em Statistics and Computing}, vol.~6, pp.~353-366.

\item
  Neal, R.~M.\ (1996) {\em Bayesian Learning for Neural Networks}, Lecture
  Notes in Statistics No.~118, New York: Springer-Verlag.

\item
  Ripley, B. D.\ (1987) {\em Stochastic Simulation}, New York: John Wiley.

\item
  Torrie, G.~M.\ and Valleau, J.~P.\ (1977) ``Nonphysical sampling distributions
  in Monte Carlo free-energy estimation: Umbrella sampling``, {\em Journal
  of Computational Physics}, vol.~23, pp.~187-199.

\end{description}
\end{document}

%% file: fig-uni1.tex
\begin{figure}[p]

\vspace*{-0.2in}

\centerline{\hbox{
\psfig{file=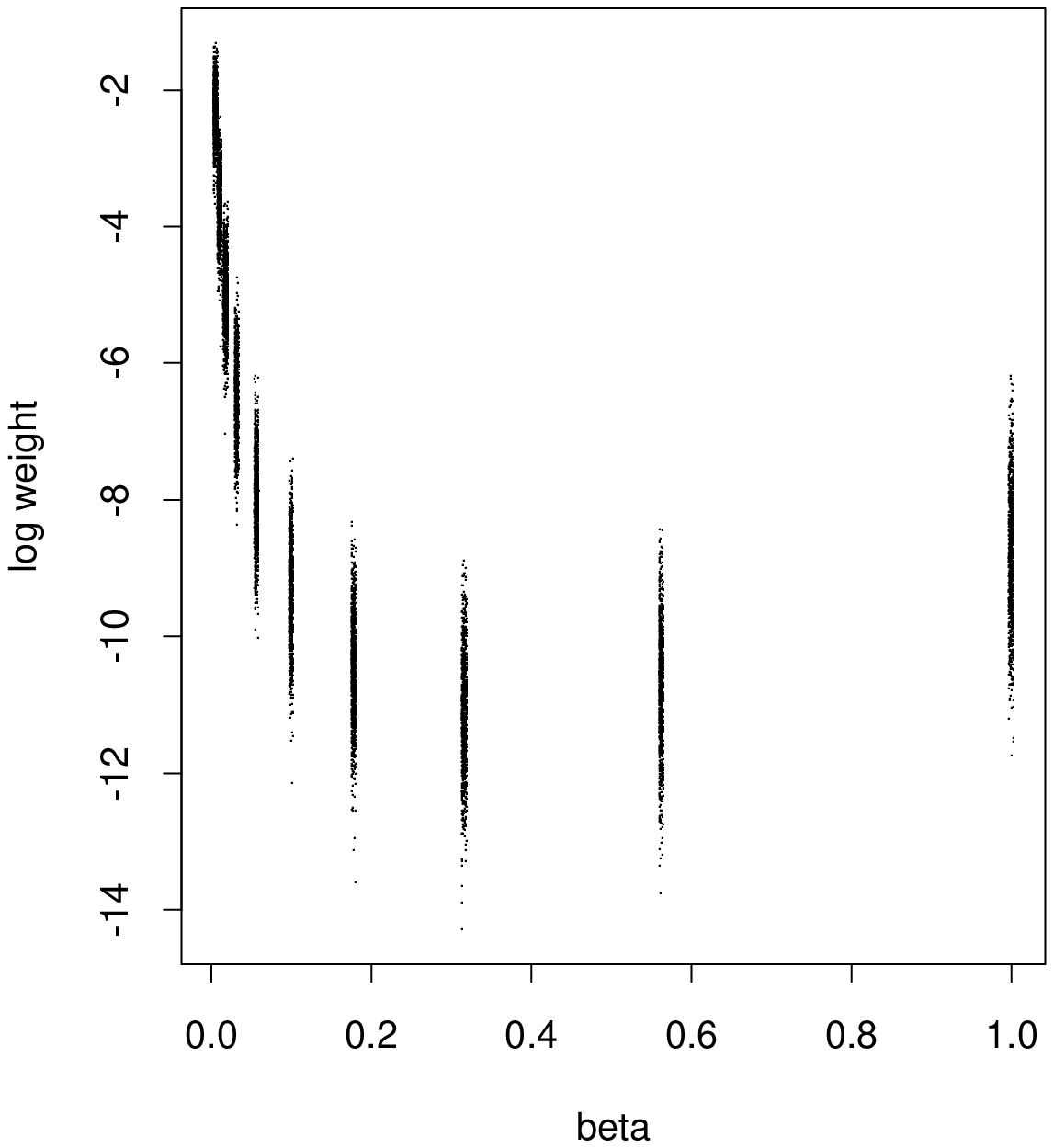,height=4in,width=3.1in}
\psfig{file=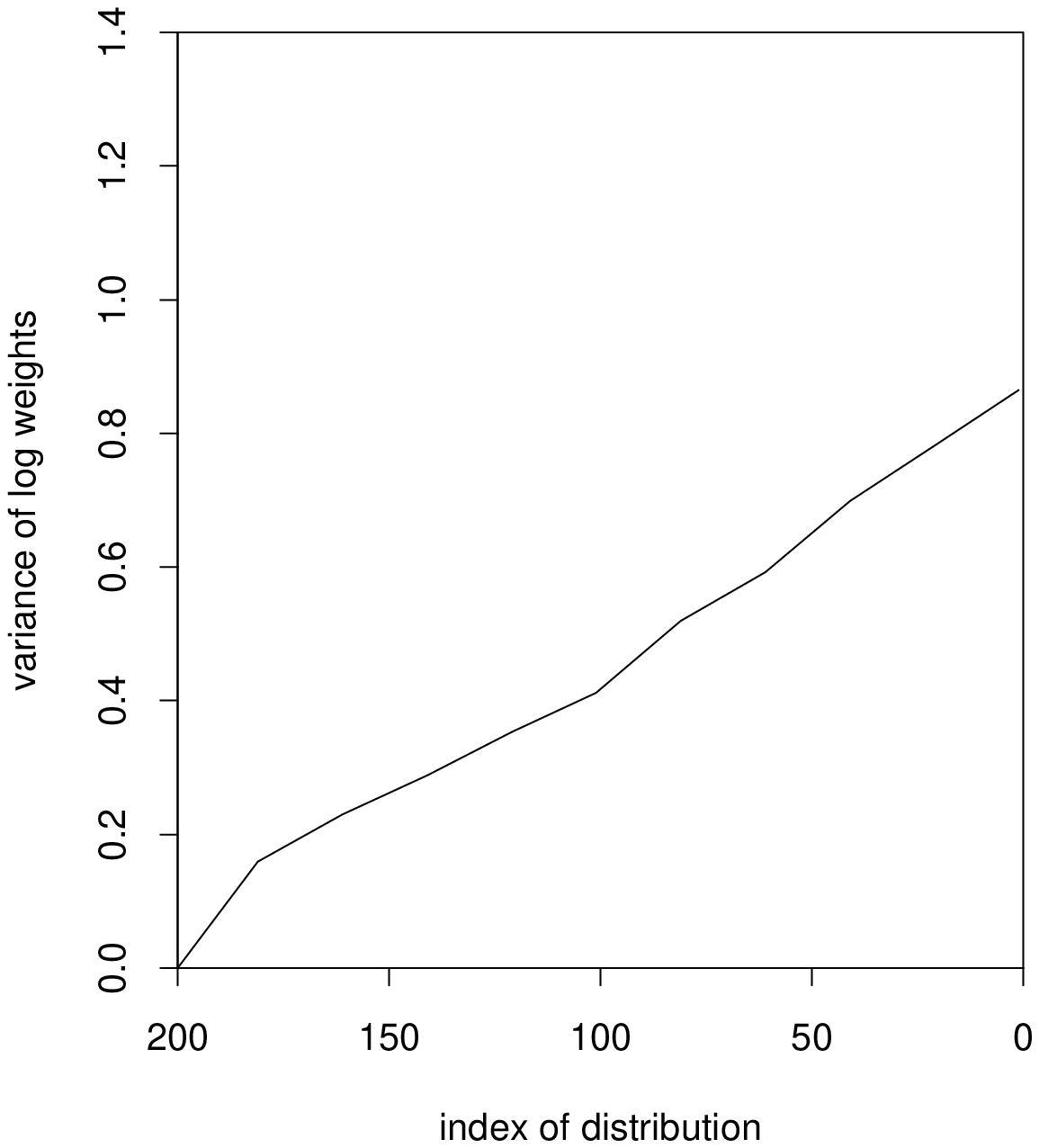,height=4in,width=3.1in}
}}

\vspace*{-0.2in}

\centerline{\hbox{
\psfig{file=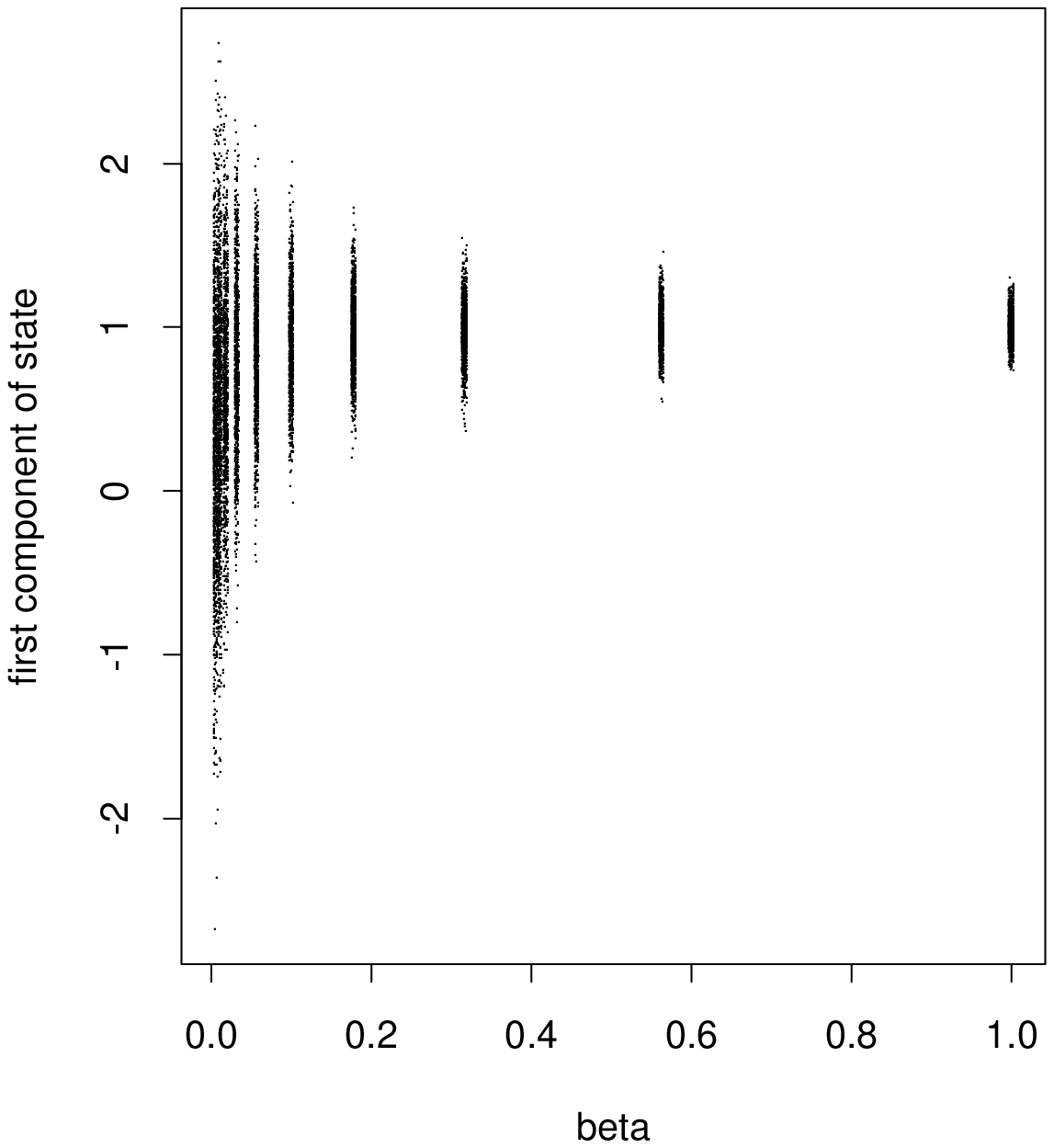,height=4in,width=3.1in}
\psfig{file=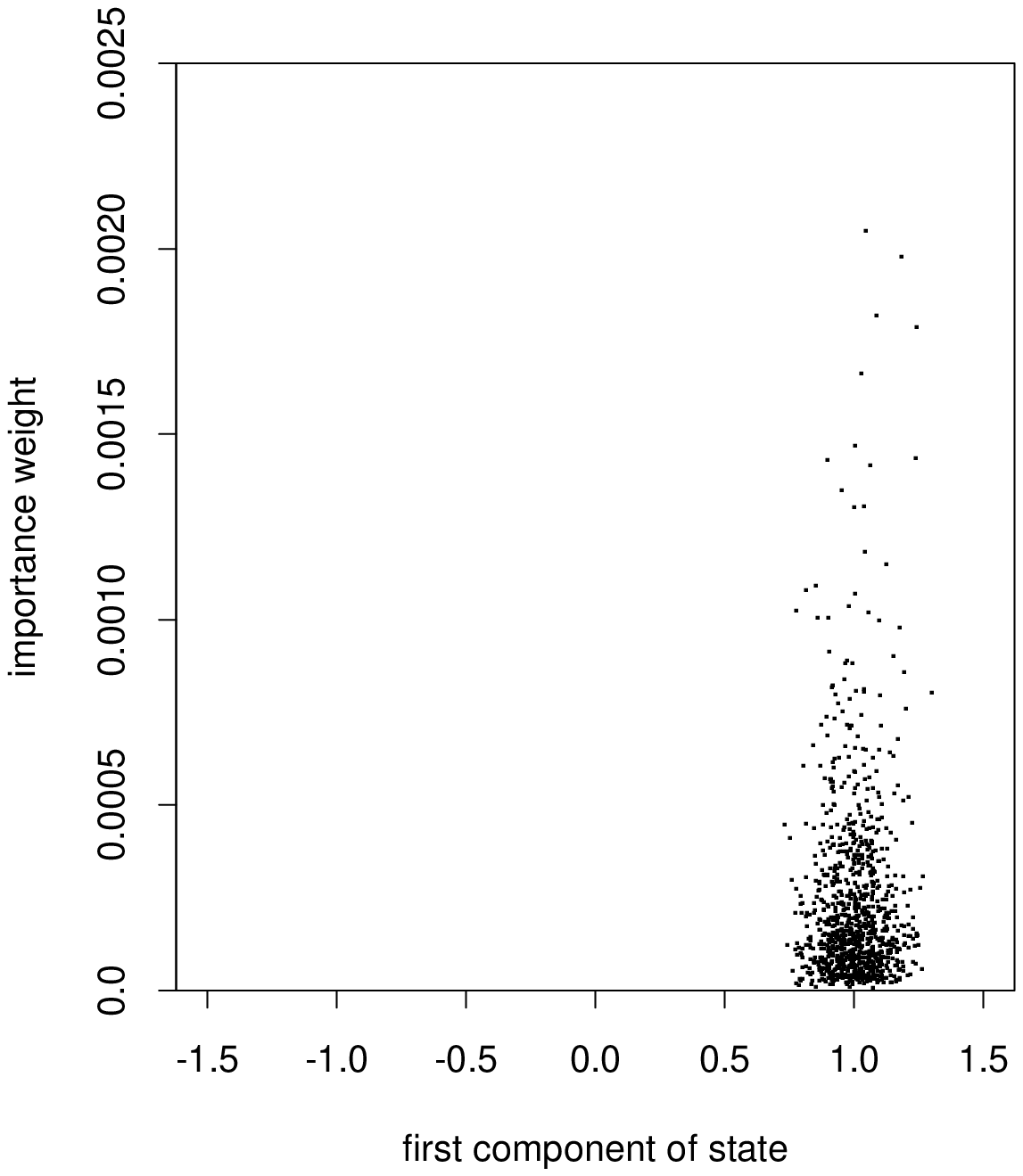,height=4in,width=3.1in}
}}

\vspace*{-0.1in}

\caption[]{Results of the first test on the unimodal distribution.
Upper left: the logs of the importance weights at ten values of
$\beta$, for each of the 1000 runs.  Upper right: the variance of the
log weights as a function of the index of $\beta$.  Lower left: the
distribution of the first component of the state at ten $\beta$
values.  Lower right: the joint distribution of the first component
and the importance weight at the ends of the runs.  Random jitter was
added to the $\beta$ values in the plots on the left to improve the
presentation.}\label{fig-uni1}

\end{figure}

%% file: fig-multi1.tex
\begin{figure}[p]

\vspace*{-0.2in}

\centerline{\hbox{
\psfig{file=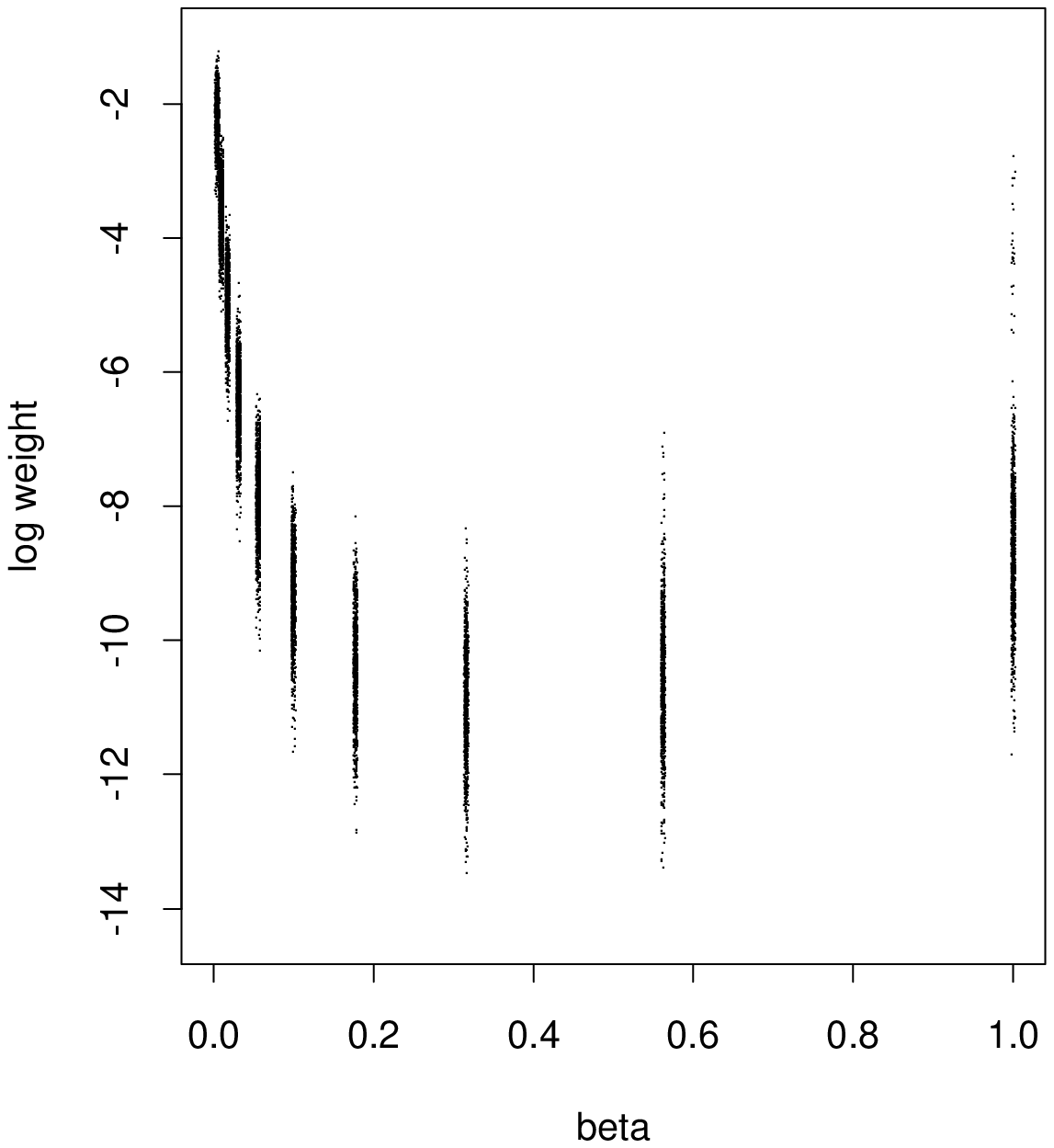,height=4in,width=3.1in}
\psfig{file=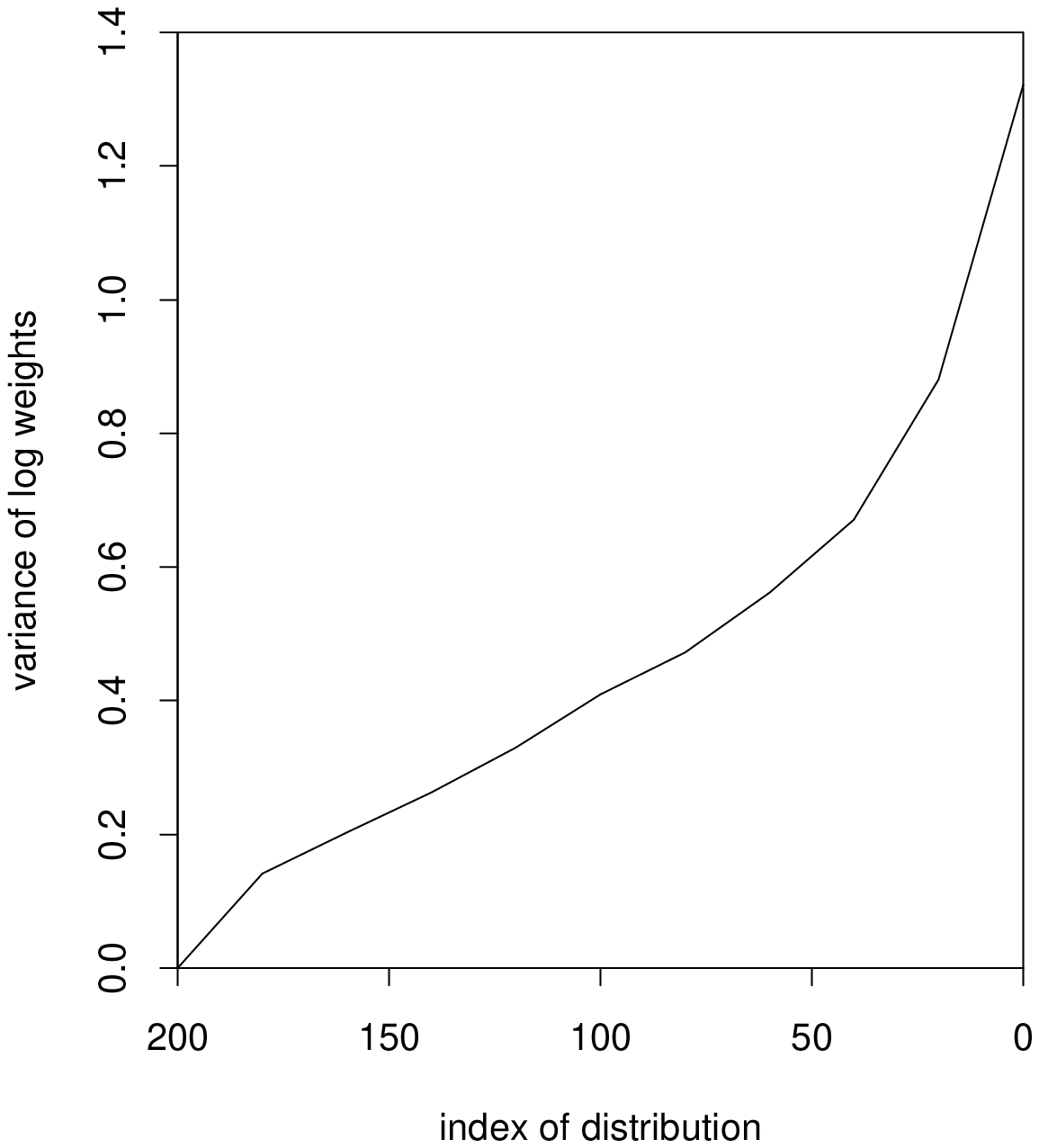,height=4in,width=3.1in}
}}

\vspace*{-0.2in}

\centerline{\hbox{
\psfig{file=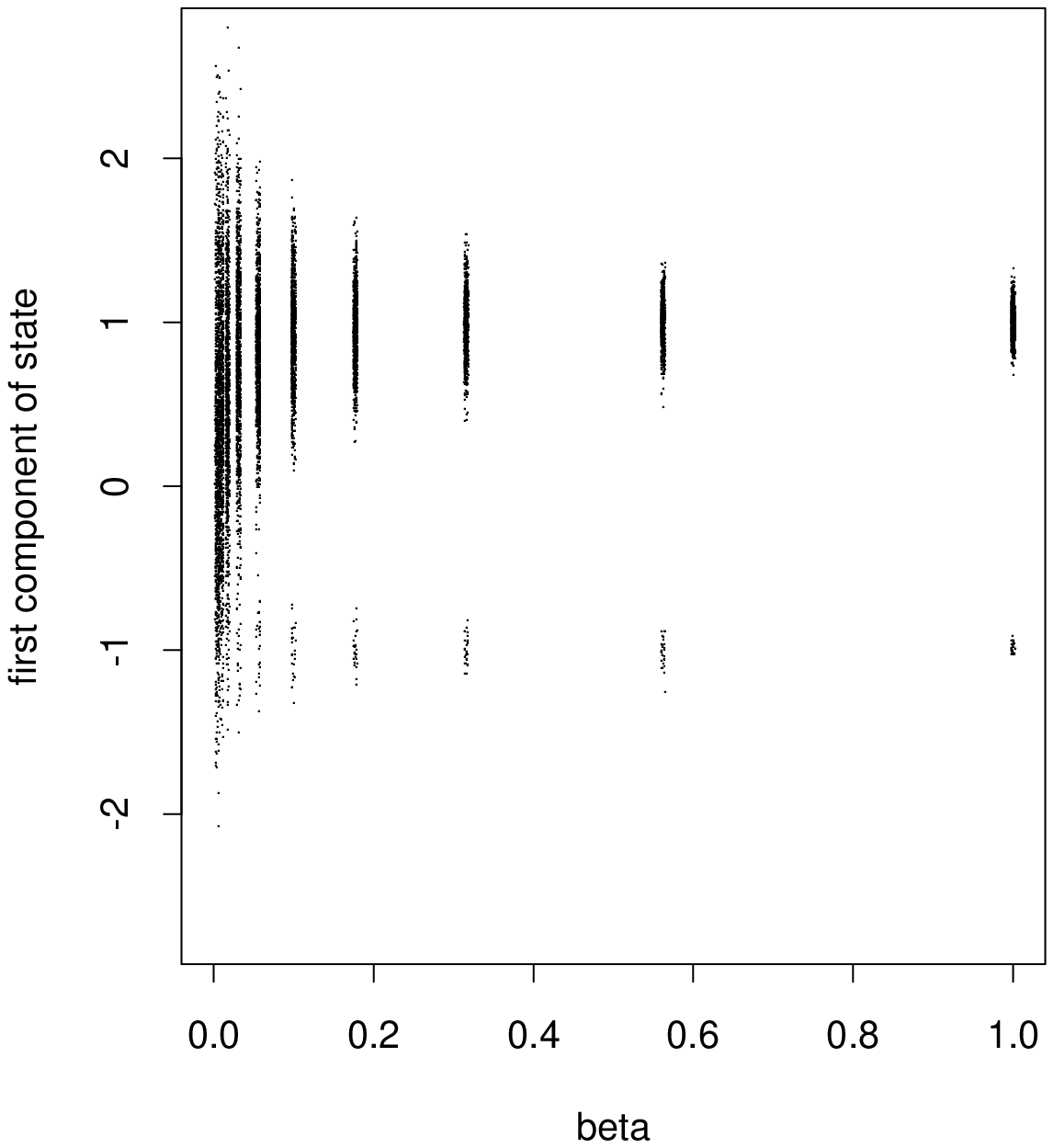,height=4in,width=3.1in}
\psfig{file=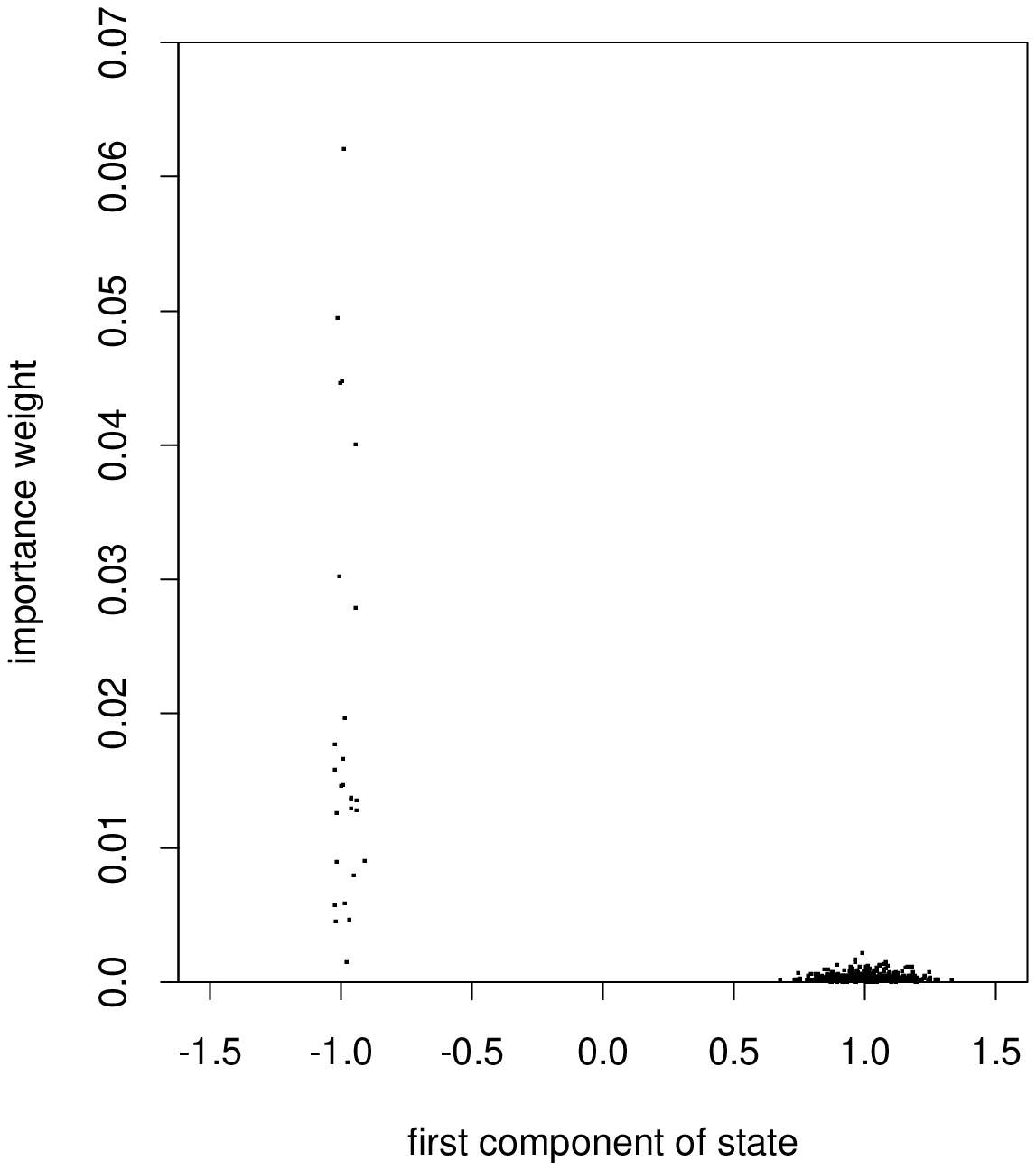,height=4in,width=3.1in}
}}

\vspace*{-0.1in}

\caption[]{Results of the first test on the distribution with two modes.
The four plots here correspond to those in Figure~\ref{fig-uni1}. \\ \\ \\ \\
}\label{fig-multi1}

\end{figure}